\documentclass[aps,twocolumn,groupedaddress]{revtex4}
\usepackage{psfig}
\usepackage{latexsym,url}
\usepackage{amsmath}

\begin{document}
\bibliographystyle{apsrev}


\title[Schweitzer/Tilch: Self-Assembling of Networks
...]{Self-Assembling of Networks \\ in an Agent-Based Model} 
\author{Frank Schweitzer}
\email[]{schweitzer@ais.fhg.de} \homepage[]{http://www.ais.fhg.de/~frank/}
\affiliation{Fraunhofer Institute for Autonomous Intelligent Systems, Schloss Birlinghoven, 53754 Sankt
    Augustin, Germany}
\affiliation{Institute of Physics, Humboldt University,
    Invalidenstra{\ss}e 110, 10115 Berlin, Germany}
\author{Benno Tilch}
\affiliation{Institute of Physics, Humboldt University,
    Invalidenstra{\ss}e 110, 10115 Berlin, Germany}




\date{April 2002}

\begin{abstract}
  We propose a model to show the self-assembling of network-like
  structures between a set of nodes without using preexisting positional
  information or long-range attraction of the nodes.  The model is based
  on Brownian agents that are capable of producing different local
  (chemical) information and respond to it in a non-linear manner. They
  solve two tasks in parallel: (i) the detection of the appropriate
  nodes, and (ii) the establishment of stable links between them.  We
  present results of computer simulations that demonstrate the emergence
  of robust network structures and investigate the connectivity of the
  network by means of both analytical estimations and computer
  simulations. 
\end{abstract}
\maketitle
\renewcommand{\bbox}[1]{\mbox{\boldmath $#1$}}
\renewcommand{\epsilon}{\varepsilon}
\newcommand{\eps}{\varepsilon}
\newcommand{\mean}[1]{\left\langle #1 \right\rangle}
\newcommand{\abs}[1]{\left| #1 \right|}
\newcommand{\eqn}[1]{eq. (\ref{#1})}
\newcommand{\Eqn}[1]{Eq. (\ref{#1})}
\newcommand{\sect}[1]{Sect. \ref{#1}}
\newcommand{\eqs}[2]{eqs. (\ref{#1}), (\ref{#2})}
\newcommand{\pic}[1]{Fig. \ref{#1}}
\newcommand{\name}[1]{{\rm #1}}
\newcommand{\bib}[4]{\bibitem{#1} {\rm #2} (#4): #3.}
\newcommand{\ul}[1]{\underline{#1}}
\newcommand{\D}{\displaystyle}
\newcommand{\pp}{\partial}

\frenchspacing

%
\section{Introduction} 

The \emph{emergence} of network structures, i.e.  the
\emph{self-organized} formation of links between a set of nodes is of
crucial importance in many different fields. In electronic engineering,
for instance, one is interested in the \emph{self-assembling} and
\emph{self-repairing} of electronic circuits
\citep{wang-01,cui-lieber-01,mange-stauffer-94}, while in biology models
for the self-wiring of \emph{neuronal networks} are investigated
\citep{segev-physlett-98, segev-00}.  On the social level, the
self-organization of human trail networks between different destinations
is a similar problem \citep{helbing-fs-et-97}.  Also the establishment of
connections on demand in telecommunication or logistics is related to the
problem discussed here.

A desirable feature of self-organized networks is their
\emph{adaptivity}. This means that new nodes can be linked to the
existing network or linked nodes can be disconnected from the network if
this is required, e.g. by the change of some external conditions.
Noteworthy, such a behavior should be not governed by a ``supervisor'' or
``dispatcher'', it should rather result from the adaptive capabilities of
the network itself.

Such problems become even more complicated if no \emph{a priori}
information about the network structure is provided, i.e.  the network
has to self-organize itself not only regarding the links but also
regarding the nodes. This is the case for instance if the nodes to be
linked to the network are ``unknown'' in the sense, that they
\emph{first} have to be \emph{discovered} and only \emph{then} can be
\emph{connected}. A common biological example is the formation of a trail
system in ants to connect a nest to a set of food sources that first have
to be found \citep{crist-haefner-94, fs-lao-family-97}. Such networks are
known to be rather flexible and adaptive.  After food sources are
exhausted, they are ``disconnected'' from the existing network, because
they are no longer visited and the respective trail is no longer
maintained, but newly found food sources can be linked to the existing
network as well.

Another important example for this kind of phenomena can be found in the
self-wiring of neural structures. A neuron that grows from the retina of
the eye towards the optic tectum (or superior colliculus) of the brain,
does not ``know'' from the outset about its destination node in the
brain, hence it has to navigate through an unknown environment in order
to detect and to reach the appropriate area.  Neural growth cones appear
to be guided by at least four different mechanisms: contact attraction,
chemoattraction, contact repulsion, and chemorepulsion
\citep{tessier-goodman-96}. These mechanisms seem to act simultaneously
and in a coordinated manner to direct pathfinding.  Once these specific
pathways are established, neuronal growth cones can navigate over long
distances to find their correct targets.

It is known that gradients of different chemical cues play a considerable
role in this navigation process. They provide a kind of \emph{positional
  information} for the navigation of the growth cones \citep{gierer-83}.
Already in 1963 Sperry \citep{sperry-63} proposed that positional
information might be encoded in the form of gradients of signaling
molecules that could be detected by the axons. I.e.  axons could read
positional information at every point on the tectum.  Noteworthy, such an
explanation assumes that the positional information resulting from the
different gradients preexists in the environment. It may then provide a
kind of long-range attraction or repulsion for the growth cones, which act
together with other short-range mechanismis.

This points to the question that shall be answered in this paper: Is it
possible to link a set of nodes without using preexisting positional
information or any kind of long-range attraction of the nodes? Can the
process of generating positional information, i.e. the detection of
``unknown'' nodes and the estabishment of chemical gradients, \emph{and}
the process of network formation, i.e. the establishment of links between
nodes, occur in parallel, on a comparable time scale, as a process of
co-evolution?

In order to show this, in Sect. II we introduce a model of Brownian
agents that are capable of producing different local (chemical)
information and respond to it in a non-linear manner. In Sect. III, this
model is applied to the formation of a network between a set of nodes. We
present results of computer simulations that demonstrate the emergence of
network structures. In Sect. IV, we investigate the network connectivity
as a particular quantitative feature of the network by means of both
analytical estimations and computer simulations. In Sect. V, we conclude
the results and comment also on the agent-based method used in this
paper.

\section{Model of Brownian Agents}

The self-organization of a network is, in the considered case, based on
two different kind of \emph{activities}: (i) the generation of positional
information in terms of chemical gradients, (ii) the nonlinear response
to the existing information in order to link the different nodes.  These
rather complex processes are not performed by usual physical particles,
therefore we have introduced the concept of \emph{Brownian agents}
\citep{fs-book-01} as a simple way to consider certain activities within
the framework of statistical physics.

A Brownian agent $i$ is characterized by different state variables that
could be either external variables such as its position $\bbox{r}_{i}$ or
its velocity $\bbox{v}_{i}$, or \emph{internal degrees of freedom}.
$\theta_{i}$ for example is assumed a discrete valued parameter that
allows to describe different responses of agent $i$ to external signals,
or different changes of its environment. Because all kind of activities
need energy, the agent's energy depot, $e_{i}$ is another important
internal degree of freedom \citep{fs-eb-tilch-98-let}.  These state
variables may change in the course of time, either or both by
deterministic and stochastic influences. Similar to the description of
Brownian motion, we will use a generalized Langevin equation for the
Brownian agent (which also justifies its denotation).  For the change of
the agent's position we may assume an overdamped Langevin equation:
\begin{equation}
\frac{d\bbox{r}_{i}}{dt}=\alpha_{i}\left.
\frac{\partial h^{e}(\bbox{r},t)}{\partial
  \bbox{r}}\right|_{\bbox{r}_{i},\theta_{i}}  
+ \sqrt{2\eps_{i}}\,\xi_{i}(t)
\label{langev-red2}
\end{equation} 
The second term denotes the stochastic influences, where $\xi_{i}(t)$ is
white noise with $\mean{\xi_{i}(t)}=0$ and
$\mean{\xi_{i}(t)\,\xi_{j}(t')} = \delta_{ij}\,\delta(t-t')$.  The
strength of the stochastic force $\eps_{i}$ could be in general an
individual parameter to weight the stochastic influences, this way it can
for example measure the individual {\em sensitivity} $\omega_{i}\propto
1/\eps_{i}$ of the agent. The first term denotes the deterministic
influences that are in the considered case assumed to result from the
\emph{gradient} of an \emph{effective field} $h^{e}(\bbox{r},t)$. This
field contains the positional information provided by different chemical
cues as specified below.  The parameter $\alpha_{i}$ describes the
strength of the individual response of the agent to the field and weights
the deterministic influences. $\alpha_{i}$ can be used to describe
different responses to the field, e.g.
\begin{itemize}
\item[(i)] attraction to the field, $\alpha_{i}>0$, or repulsion,
  $\alpha_{i}<0$
\item[(ii)] response only if the local value of the field is above a
  certain threshold $h_{0}$: $\alpha_{i}=\Theta[h^{e}(\bbox{r},t)-h_{0}]$,
  with $\Theta[y]$ being the Heavyside function: $\Theta=1$, if $y>0$,
  otherwise $\Theta=0$.
\item[(iii)] response only if the agent has a specific internal value
  $\theta$: $\alpha_{i}=\delta_{\theta_{i},\theta}$. 
\end{itemize}
Throughout this paper, we
assume the two individual parameters as constants: $\alpha_{i}\equiv
\alpha=1$ and $\eps_{i}=D_{n}$ where $D_{n}$ is the spatial diffusion
coefficient, but we want to mention that the idea of an adjustable
sensitivity has been successfully applied to model search problems with
Brownian agents \citep{fs-lao-family-97,fs-97-agent}.

In addition to the movement of the Brownian agents, i.e. changes of their
state variables $\bbox{r}_{i}$, we also have to consider changes of their
internal degree of freedom $\theta_i(t)$ that in this application should
have one of the following values: $\theta_{i} \in \{0,-1,+1\}$.
Initially, $\theta_i(t_0)=0$ holds for every agent. The parameter
$\theta_i$ can be changed in the course of time by an interaction between
the moving agents and the nodes. To be specific, we consider a
two-dimensional surface, where a number of $j=1,...,z$ nodes are located
at the positions $\bbox{r}^{z}_{j}$ (cf.  \pic{raut0}). A number of $z_+$
nodes should be characterized by a positive potential, $V_j=+1$, while
$z_-=z-z_+$ nodes have a negative potential, $V_j=-1$.  We note
explicitely, that the nodes do \emph{not} have any \emph{long-range
  effect} on the agents, such as attraction or repulsion. Their effect is
restricted to their location, $\bbox{r}^{z}_{j}$.
\begin{figure}[ht]
\centerline{\psfig{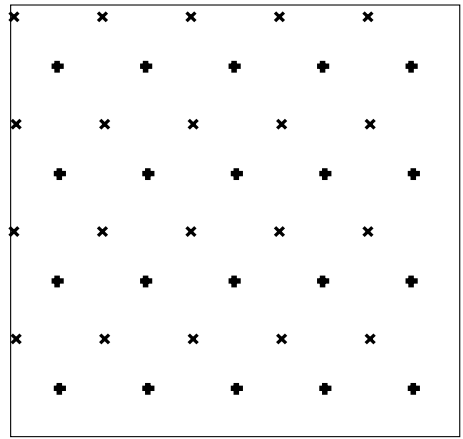}}
\caption[p-raut0]{
  Example of a regular distribution of 40 nodes on a $100 \times 100$
  lattice. For the computer simulations, periodic boundary conditions
  have been used. $z_{+}=20$, $z_{-}=20$. \{x\} indicates nodes with a
  potential $V_{j}=-1$, \{+\} indicates nodes with a potential
  $V_{j}=+1$. \label{raut0}}
\end{figure}

It is the (twofold) task of the  Brownian agents, first to
\emph{discover} the nodes and then to \emph{link} nodes with an opposite
potential, this way forming a self-organized network between the set of
nodes. If an agent hits one of the nodes, its internal degree of freedom
is changed due to the following equation: 
\begin{equation}
\Delta \theta_i(t)= 
\sum_{j=1}^z (V_j-\theta_i)\;\frac{1}{A} \int_{A}
\,\delta\Big(\bbox{r}^{z}_{j}-\bbox{r}_{i}(t)\Big) \,d\bbox{r} 
\label{change}
\end{equation}
The delta function is equal to $1$ only for
$\bbox{r}^{z}_{j}=\bbox{r}_{i}$ and zero otherwise. So, eq.
(\ref{change}) indicates, that a agent changes its internal state,
$\theta_i$, only if it hits one of the nodes. Then it takes over the
value of the potential of the respecting node, $V_j$, which means
$\theta_i$ remains constant if $V_j=\theta_i$, and $\theta_i \rightarrow
V_j$, if $V_j \neq \theta_i$.  We note that the probability for a
(pointlike) agent to hit a (pointlike) node is almost vanishing. However,
the computer simulations discussed in the following section are carried
out on a discrete lattice, so the agent and the node both have a finite
extension, in which case \eqn{change} makes sense.

If the Brownian agent hits one of the nodes, this impact may result in an
active state of the agent - a {\em kick}, originated by the potential,
which may change the internal parameter, $\theta_i$, due to eq.
(\ref{change}). In the active state, it is assumed that the agent is able
to produce a chemical, either component $(-1)$ or $(+1)$, in dependence
on the actual value of the internal parameter. We note that the agent's
ability to produce the chemical in general depends on another internal
parameter, namely the internal energy depot that may set limits to the
agent's activities. In this model however it is assumed that the internal
energy depot is always sufficiently balanced, thus its influence shall be
neglected here.  The agent's chemical production rate, $s_i(\theta_i,t)$,
is assumed as follows:
\begin{equation}
  \begin{aligned}
s_i(\theta_i,t)=\frac{\theta_i}{2}\Big[&(1+\theta_i)\,s^{0}_{+1}\,
\exp\{-\beta_{+1}\,(t-t_{n+}^{i})\}\\  -\,&(1-\theta_i)\,s^{0}_{-1}\,
\exp\{-\beta_{-1}\,(t-t_{n-}^{i})\}\Big]
\label{prod}
\end{aligned}
\end{equation}
Eq. (\ref{prod}) means that the agent is not active, as long as
$\theta_i=0$, which means before it hits one of the nodes the first time.
After that event, the agent begins to produce either component $(+1)$ if
$\theta_i=+1$, or component $(-1)$ if $\theta_i=-1$. This activity,
however, goes down with time, expressed in an exponential decrease of the
production rate. Here, $s^{0}_{+1}, s^{0}_{-1}$ are the initial
production rates and $\beta_{+1}, \beta_{-1}$ are the decay parameters
for the production of the chemical components $(+1)$ or $(-1)$.
Respectively, $t^{i}_{n+}, t^{i}_{n-}$ are the times, when the agent
$i$ hits either a node with a positive or a negative potential.

The spatio-temporal concentration of the chemicals shall be described by
a \emph{chemical field} $h_{\theta}(\bbox{r},t)$ consisting either of
component $(+1)$ or $(-1)$, which obeys the following equation:
\begin{equation}
\frac{\pp h_{\theta}(\bbox{r},t)}{\pp t}=-k_{\theta}\,
h_{\theta}(\bbox{r},t) + \sum_{i=1}^{N} 
s_i(\theta_i,t)\;\delta_{\theta;\theta_{i}}\;
\delta\Big(\bbox{r}-\bbox{r}_i(t)\Big) 
\label{h-net-nd}
\end{equation}
The first term describes the exponential decay of the existing
concentration due to spontaneous decomposition of the chemical, where
$k_{\theta}$ is the decomposition rate. The second term denotes the
production of the field by the agents.  Here,
$\delta_{\theta;\theta_{i}}$ means the Kronecker Delta used for discrete
variables, indicating that the agents only contribute to the field
component that matches their internal parameter $\theta_{i}$. The Delta
function $\delta(\bbox{r}-\bbox{r}_i(t))$ means that the agents
contribute to the field only \emph{locally}, at their current position,
$\bbox{r}_{i}$.  Diffusion of the chemical substances is not
considered here.

The \emph{effective} field, $h^{e}(\bbox{r},t)$, is a specific function
of the different components of the field, \eqn{h-net-nd}.  It should
influence the movement of the agents according to the overdamped Langevin
\eqn{langev-red2} and dependent on their current internal parameter,
$\theta_{i}$, as follows:
\begin{equation}
    \frac{\partial h^{e}(\bbox{r},t)}{\partial \bbox{r}} \!=\!
    \frac{\theta_i}{2}\!\left[ (1+\theta_i)\frac{\partial
        h_{-1}(\bbox{r},t)}{\partial \bbox{r}}-
      (1-\theta_i)\frac{\partial h_{+1}(\bbox{r},t)}{\partial
        \bbox{r}}\right]
\label{dh-eff}
\end{equation}
\Eqn{dh-thet} summarize the non-linear feedback between the field and the
agents, as given by the eqs. (\ref{prod}), (\ref{h-net-nd}),
(\ref{dh-eff}):
\begin{equation}
  \label{dh-thet}
  \begin{array}{ccc}
\quad \theta_{i} \quad & \quad \bbox{\nabla}_{i}\,h^{e}(\bbox{r},t) \quad 
&  \quad s_{i}(\theta_{i},t) \quad \\ \hline
0 & 0 & 0 \\
+1 & \bbox{\nabla}_{i}\,h_{-1}(\bbox{r},t) &
 s_{i}(+1,t)\\
-1 & \bbox{\nabla}_{i}\,h_{+1}(\bbox{r},t)&
s_{i}(-1,t)
\end{array}
\end{equation}
Before presenting computer simulations, we would like to summarize our
model of network formation that is introduced here in terms of a
\emph{agent-based} approach. Each agent is active in the sense that it
can (i) move, (ii) produce locally one out of two different chemical cues
and (iii) respond to local gradients of these different chemicals.  The
actions of all agents are coupled indirectly via an effective field that
is comprised of the two different chemical components.  The agent
activities further depend on an internal parameter $\theta_{i}$ that
allows to describe a different ``behavior'': Our model assumes, that
agents with an internal state $\theta_i=0$ do not contribute to the field
and are not affected by the field. They simply move like Brownian
particles. Agents with an internal state $\theta_i=+1$ contribute to the
field by producing the chemical cue $(+1)$, while they are affected by the
part of the field that is determined by chemical $(-1)$. On the other hand,
agents with an internal state $\theta_i=-1$ contribute to the field by
producing chemical $(-1)$ and are affected by the part of the field, which
is determined by component $(+1)$. Moreover, if the agent hits one of the
nodes, the internal state can be switched due to eq. (\ref{change}).
Hence, the agent begins to produce a different chemical while beeing
affected by the opposite potential. Precisely, at one time the agent does
\emph{not} respond to the gradient of the same field component, which it
contributes to via producing a chemical.

As the result of this non-linear feedback between the  Brownian
agents and the effective field generated by them, we can observe the
formation of macroscopic structures shown in the following section. 

\section{Simulation Results of Network Formation}

For the computer simulations, a triangular lattice with periodic boundary
conditions was used.  Further, we have assumed that the parameters
describing the production and decay of the chemical, are the same for
both components:
\begin{equation}
  \label{convent}
s^{0}_{+1} = s^{0}_{-1} = s_{0};\;
k_{+1} = k_{-1} = k_{h};\;
\beta_{+1} = \beta_{-1} = \beta
\end{equation}
The agents start initially at random positions and with the internal
parameter $\theta_{i}(t=0)=0$.  For the evolution of the network, we
evaluate the sum $\hat{h}(\bbox{r},t)$ of the two field components
generated by the agents.  For the plots, however, we have to match these
values with a \emph{grey scale} of 256 values, which is defined as
follows:
\begin{eqnarray}
  \label{grey}
c(\bbox{r},t) &=& 255 \left[1- \log\left(1 + 9\,
\frac{\hat{h}(\bbox{r},t)-\hat{h}_{min}(t)}{\hat{h}_{max}(t)
-\hat{h}_{min}(t)}\right)\right] \nonumber \\
\hat{h}(\bbox{r},t)&=&h_{+1}(\bbox{r},t)+h_{-1}(\bbox{r},t)
\end{eqnarray}
This means that the highest actual value, $\hat{h}_{max}(t)$, always
refers to \emph{black} ($c=0$ in PostScript), whereas the actual minimum
value, $\hat{h}_{min}(t)$ encodes \emph{white} ($c=255$). Both extreme
values change of course in time, therefore each snapshot of the time
series presented has its own value mapping.

As a first example, we show the evolution of the connections between four
nodes (\pic{karo}).  In the course of time agents that have by chance
discovered a node (this way going over into an active state) begin to
perform a \emph{directed motion} between the different nodes. Eventually,
a link appears which can be clearly distinguished from the surrounding.
\begin{figure}[htbp]
\centerline{
\hfill \psfig{figure=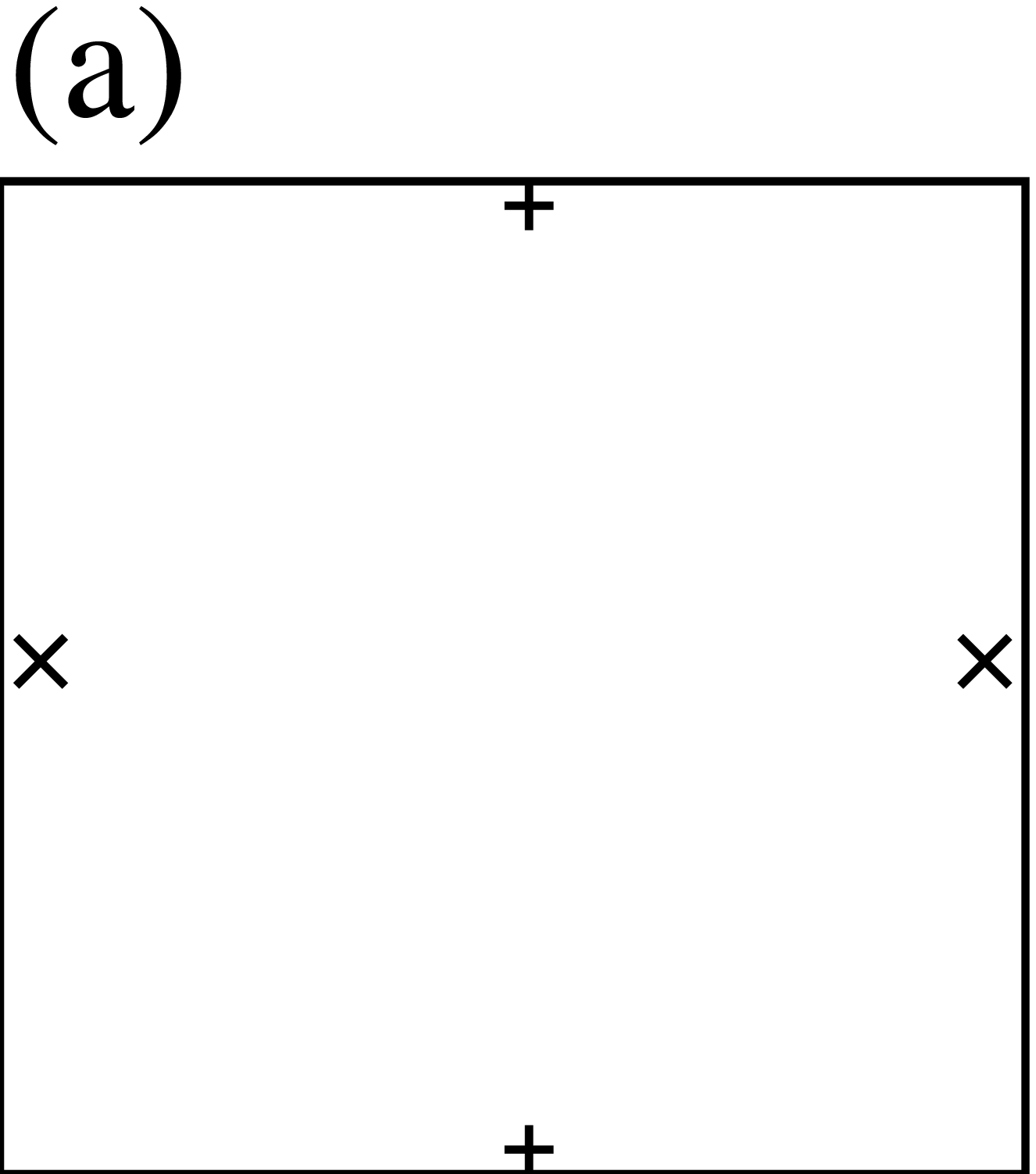,width=3.5cm}
\hfill
\psfig{figure=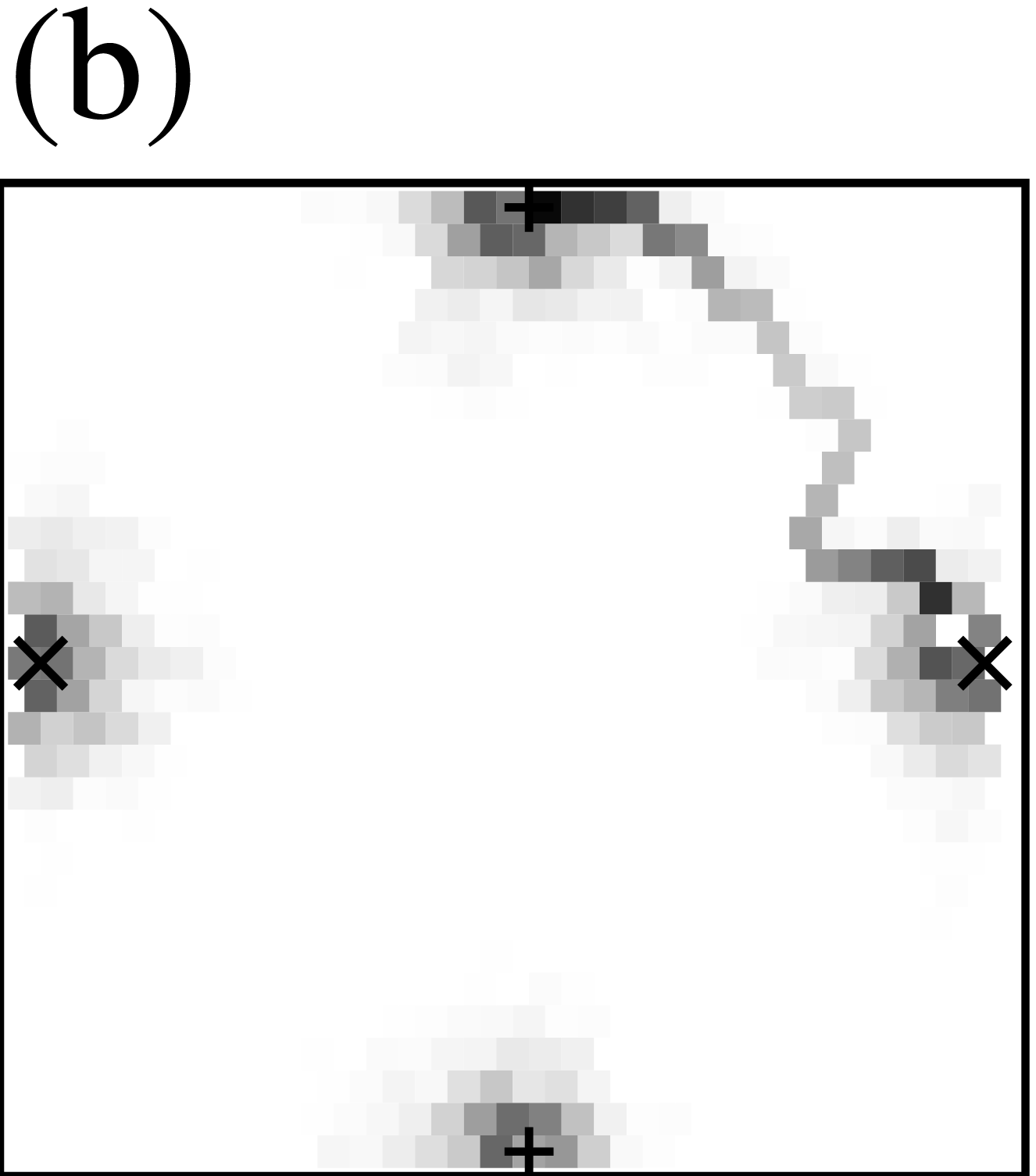,width=3.5cm}\hfill
}\vspace*{0.8cm}
\centerline{\hfill
\psfig{figure=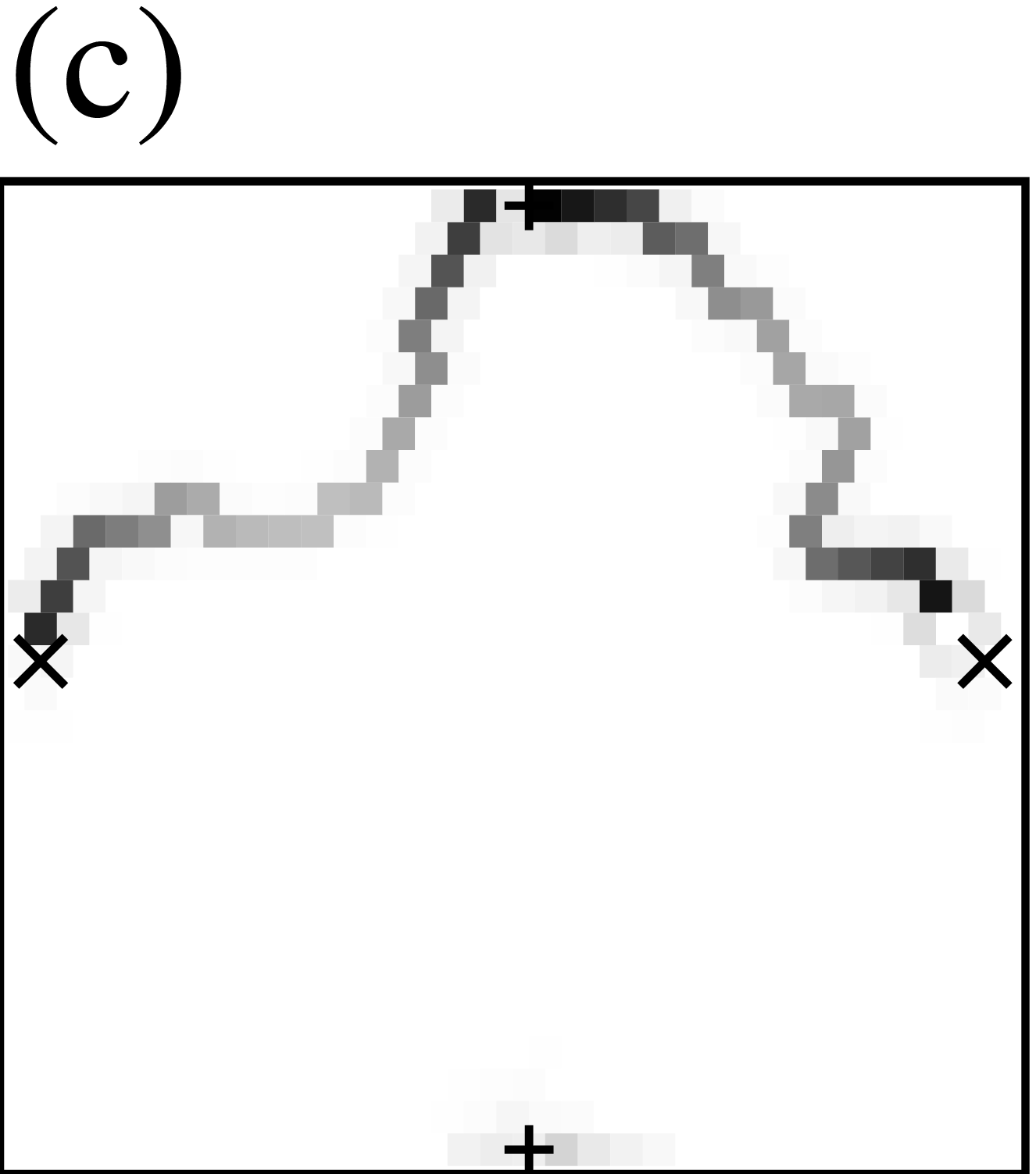,width=3.5cm}
\hfill
\psfig{figure=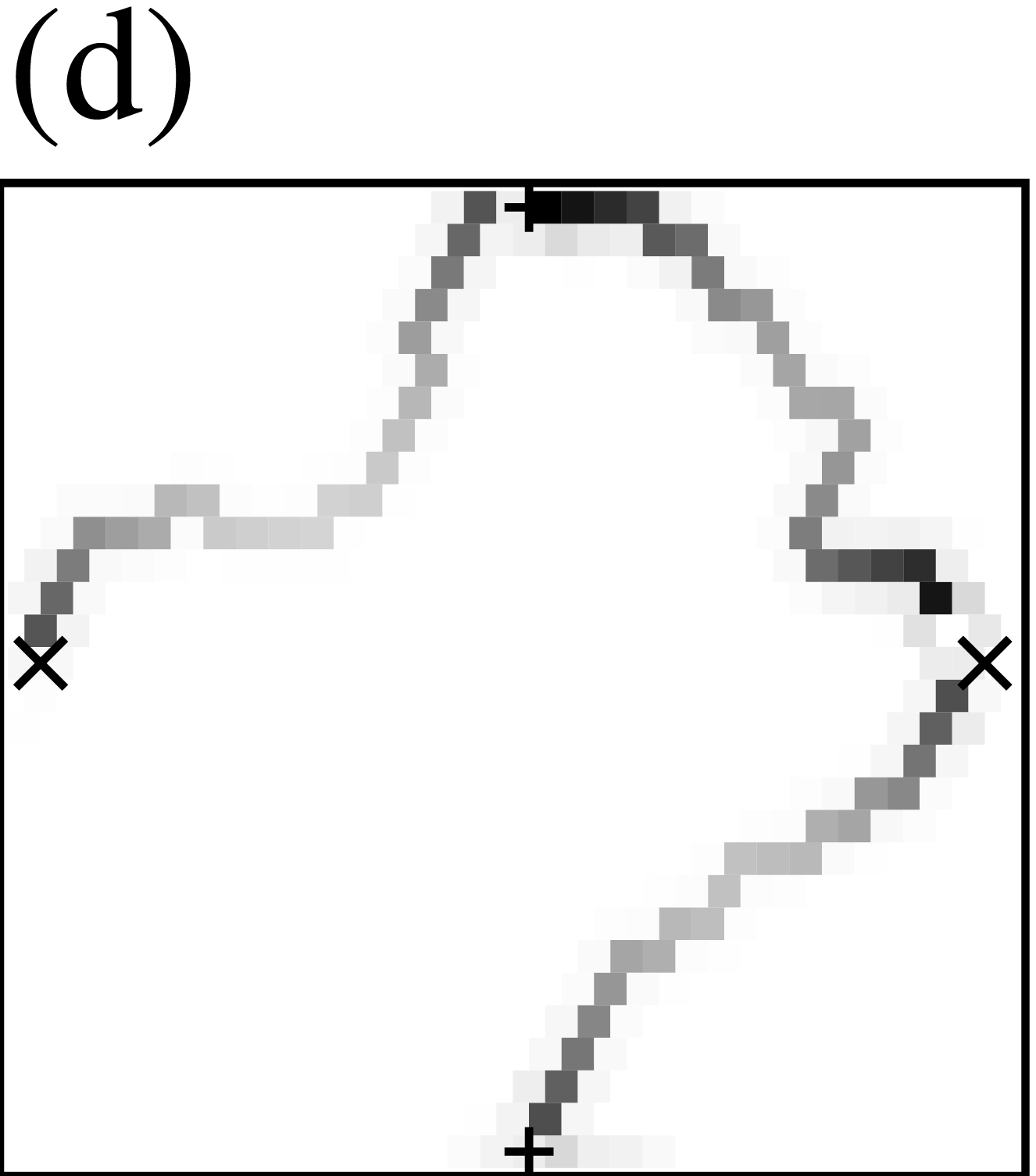,width=3.5cm}\hfill}
\caption[figx]{
  Formation of links between 4 nodes: (a) initial state, (b) after 100
  simulation steps, (c) after 1.000 simulation steps, (d) after 4.500
  simulation steps. Lattice size $30 \times 30$, 450 agents.  Parameters:
  $s_{0}$ = 25.000, $k_{h}$ = 0.01, $\beta$ = 0.2, $k_{h}$ = 0.01.
\label{karo}}
\end{figure}

\pic{karo} would suggest that in the course of time all nodes with an
opposite potential should be connected. This however is not the case
because the existing connections cause a \emph{screening effect} that
forces the agents to move along existing connections rather than making
up new ones. This screening effect becomes more obvious, when the number
of nodes is increased. \pic{t-rautpic} shows the time evolution of a
network, which should connect 40 nodes (cf. also \pic{raut0}). We see
\footnote{A video of these computer simulations can be found at
  \texttt{http://www.ais.fhg.de/\~{}frank/network.html}.} that in the
course of time the agents aggregate along the connections, which results
in higher agent concentrations and in higher fields along the
connections.  The self-assembling network is created very fast and
remains stable in the
long run. 

\begin{figure}[htbp]
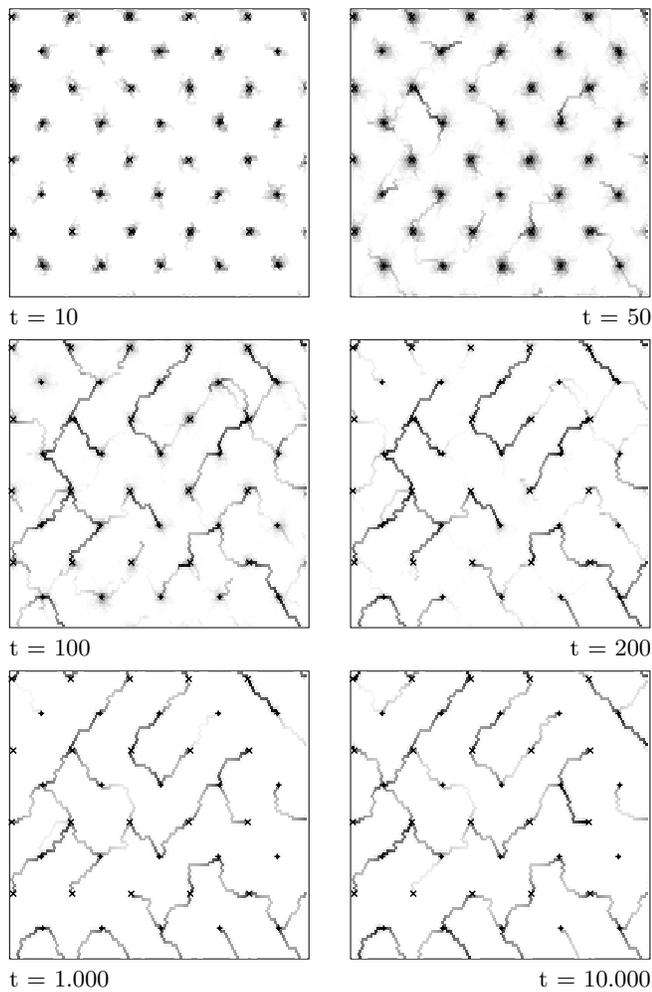

\centerline{
\psfig{figure=fig-03a.eps,height=4.0cm,width=4.0cm}\hfill
\psfig{figure=fig-03b.eps,height=4.0cm,width=4.0cm}}
\centerline{ t = 10 \hfill t = 50}
\centerline{
\psfig{figure=fig-03c.eps,height=4.0cm,width=4.0cm}\hfill
\psfig{figure=fig-03d.eps,height=4.0cm,width=4.0cm}}
\centerline{ t = 100 \hfill t = 200}
\centerline{
\psfig{figure=fig-03e.eps,height=4.0cm,width=4.0cm}\hfill
\psfig{figure=fig-03f.eps,height=4.0cm,width=4.0cm}}
\centerline{ t = 1.000 \hfill t = 10.000}
\caption{
  Time series of the evolution of a network (time in simulation steps).
  The initial state is shown in \pic{raut0}.  Parameters: $5.000$ agents,
  $s_0 = 10.000$, $k_{h}= 0.03$, $\beta = 0.2$.
\label{t-rautpic}}
\end{figure}
The time series of \pic{t-rautpic} indicates that for the formation of
the network a \emph{transient stage} exists, during which new nodes are
discovered and new connections appear. After the transient time $t_{tr}$
however the existing links are only stabilized, with small possible
fluctuations.  In order to get an estimate of the transient time, we have
evaluated the total fraction $x_{\theta}(t)=N_{\theta}(t)/N$ of agents
that currently have the internal parameter $\theta$.  The result shown in
\pic{tet-n3} is based on the simulations of \pic{t-rautpic}. It indicates
that after $t\approx 1500$ simulation steps every agent has found at
least one node by chance, thus changing its internal parameter either to
$(+1)$ or to $(-1)$.  Further, after this time the share between these
internal parameters is almost equally balanced, with slight fluctuations
around $x_{\theta}=0.5$, dependent on the actual position of the agents.

\begin{figure}[htbp]
\centerline{\psfig{figure=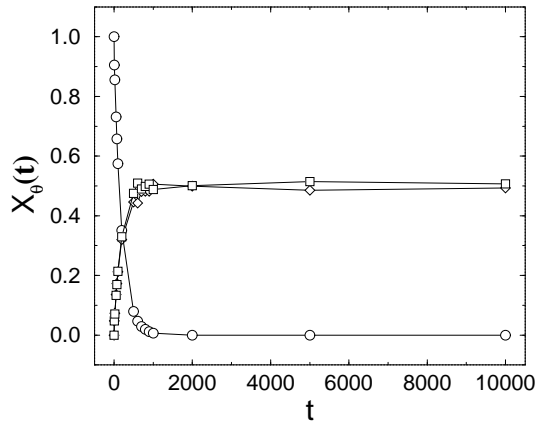,width=7.cm}}
\caption{
  Fraction $x_{\theta}$ of agents with the internal parameter $\theta$
  vs. time in simulation steps. ($\bigcirc$): $\theta=0$, ($\Diamond$):
  $\theta=+1$, ($\Box$): $\theta=-1$. The values are obtained from the
  simulation, \pic{t-rautpic}.  Initial conditions: $x_{0}=1,
  x_{\pm1}=0$. \label{tet-n3}}
\end{figure}
In \citep{fs-tilch-02-b}, we have shown that the equations for
$x_{\theta}(t)$ read explicitely:
\begin{eqnarray}
  \label{x0}
 x_{0}(t)&=& 1\, \exp\left\{ - D_{n} \frac{z_{+}+z_{-}}{A}\, t\right\}
\\
x_{\theta}(t)&=& \frac{z_{\theta}}{z_{+}+z_{-}}\,\left(1\,-\, 
\exp\left\{ - D_{n} \frac{z_{+}+z_{-}}{A}\, t\right\}\right) 
\nonumber \\
&& \theta \in \{-1,+1\} \nonumber
\end{eqnarray}
In the asymptotic limit, the fraction $x_{\theta}$ is determined by the
appropriate number of nodes, which change the internal state of the
agents into $\theta$. The transient time $t_{tr}$ can be estimated by
assuming that the difference between $x_{\theta}(t)$ and the stationary
value $x_{\theta}^{stat}$ should be smaller than a certain value,
$\kappa$:
\begin{equation}
  \label{t-trans}
  t_{tr} \geq \frac{A}{D_{n}\,z}\, \ln\left(\frac{1}{\kappa}\right)
\end{equation}
For $D_{n}=1$, we find for $\kappa=10^{-2}$ a transient time of
$t_{tr}=1150$, and for $\kappa=10^{-3}$ $t_{tr}=1700$, which is in good
agreement with the results of the computer simulations. After that time,
the assembled network should remain almost stable.

Patterns like the network shown are intrinsically determined by the
history of their creation. It means that irreversibility and early
symmetry breaks play a considerable role in the determination of the
final structure.  The location of the different nodes acts more or less
as a boundary condition for the structure formation which sets limits to
the achievable structures, but does not determine the way of connecting
the different nodes.

Despite the fact, that in \pic{t-rautpic} almost all nodes are connected
by at least one link, only some out of all possible connections have been
realized. In particular, only nearest neighbour nodes with opposite
potentials are connected. This is partly due to the screening effect that
makes longer connections an unlikely event, but also indicates that a
maximum distance $L^{\star}$ between two nodes exists, which can only be
connected by the agents. An approximation for this critical distance is
given in \citep{fs-tilch-02-c}.

Eventually, we note that the network formation is not restricted to
regular or symmetric distributions of nodes. \pic{cono} shows a
simulation, where different nodes are connected with a center.

\begin{figure}[htbp]
  \centerline{\psfig{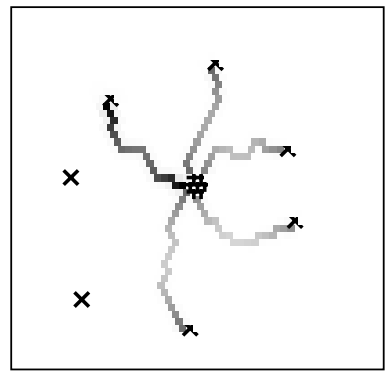}}
\caption{
  Formation of links between a center ($z_{-}=1$) and surrounding nodes
  ($z_{+}=7$) after $10.000$ simulation steps. Lattice size: $50 \times
  50$, $2.000$ agents. Parameters: $s_{0}=20.000$, $k_{h}=0.02$, $\beta =
  0.2$.
\label{cono}}
\end{figure}
In \citep{fs-lao-family-97} we have also discussed a different variant of
the model to demonstrate its flexibility in connecting additional nodes
to the network, or disconnecting obsolete ones. Further, in
\citep{fs-tilch-02-c} we have shown that the switching behavior between a
connected and a disconneced state can be very short, which would allow
the construction of a dynamic switch.

\section{Estimation of the Network Connectivity}
\subsection{Definition of Connectivity}

In order to characterize a network, one of the most important questions
is whether two nodes $k$ and $l$ are connected or not. In the model
considered a connection is defined in terms of the chemical field
$\hat{h}(\bbox{r},t)$ produced by the agents. During the first stage of
the network formation, the agents have randomly visited almost every
lattice site before their motion turned into a bound motion between the
nodes. Therefore, the field $\hat{h}(\bbox{r},t)$ has a non-zero value
for almost every $\bbox{r}$, which exponentially decays, but never
vanishes. Hence, in order to \emph{define a connection} in terms of
$\hat{h}(\bbox{r},t)$, we have to introduce a \emph{threshold value}
$h_{thr}$, which is the minimum value considered for a link. More
precisely, a \emph{connection} between two nodes $k$ and $l$ should only
exist if there is a path $a \in A$ between $k$ and $l$ along which the
actual value of the field is larger than the threshold value:
\begin{equation}
  \label{thresh-n}
 \hat{h}(a,t)>h_{thr} \;\;\; \mbox{for $a \in A$} 
\end{equation}
Such a definition does not necessarely assumes that the connection has to
be a \emph{direct link}. Instead, it could be any path $a$, which may
also include other nodes, as long as the value $\hat{h}(a,t)$ along the
path is above the threshold.

We want to define the \emph{local connectivity} $E_{lk}$ as follows:
\begin{equation}
  \label{connect}
 E_{lk} =   \left \{
 \begin{array}{ll}
1 &  \mbox{if $k$ and $l$ are connected by a path $a \in A$,} \\
 & \mbox{along which $\hat{h}(a,t)>h_{thr}$}  \\
0 & \mbox{otherwise}
\end{array} \right. 
\end{equation}
We note that the connectivity $E_{lk}$ does not change if two nodes $k$
and $l$ are connected by more than one path.

If we consider a number of $z$ nodes, then the \emph{global connectivity}
$E$ that refers to the whole network is defined as follows:
\begin{equation}
  \label{connec-e}
  E=\frac{\sum\limits_{k=1}^{z}\,\sum\limits_{l>k}^{z}\, E_{lk}}{
\sum\limits_{k=1}^{z}\,\sum\limits_{l>k}^{z}\,1}\;=\;
\frac{2}{z(z-1)}\,\sum\limits_{k=1}^{z}\,\sum\limits_{l>k}^{z}\, E_{lk}
\end{equation}
Dependent on the configuration of nodes, there may be numerous different
realizations for the connections, which result in the same connectivity
$E$. 

\subsection{Estimation of the Theshold Value}
In order to use the definition for the connectivity to evaluate the
simulated networks, we first have to define the threshold value
$h_{thr}$. This should be the \emph{minimum value} of
$\hat{h}(\bbox{r},t)$ along a \emph{stable connection} between two nodes.
For our estimations, we treat the connection between two nearest neighbor
nodes $k$ and $l$ as an \emph{one-dimensional} structure, where $x$ is
now the space coordinate, and $L$ the linear distance between the two
nodes $k$ and $l$. The node at $x=0$ should have a positive potential
$V=+1$, while the node at $x=L$ has a negative potential $V=-1$:
\begin{equation}
  \label{one-d}
  0 \leq x \leq L\;;\;\;V(0)= +1 \;;\;\;V(L) = -1
\end{equation}
We assume that a stable connection exists if both field components
$h_{\theta}(x,t)$ have reached their stationary values:
\begin{equation}
  \label{hx-stat}
  \frac{\partial h_{\theta}(x,t)}{\partial t} = - k_{\theta}\,
  h_{\theta}(x,t)+ \sum\limits_{i} s_{i}(\theta,t)\, \delta(x-x_{i})
  \,= \, 0 
\end{equation}
Of course, we do not know, how many agents are actually on the
connection between $k$ and $l$.  For our estimations we have to bear in
mind that $h_{thr}$ should determine the \emph{lower limit} of the
possible values of $\hat{h}$, therefore it is justified to assume the
worst case, which means that the local number of agents at a specific
location is just given by the \emph{average agent density}
$\bar{n}=N/A$, where $N$ is the total agent number and $A$ is the
surface size. Further, we found in the computer simulations (cf.
\pic{tet-n3}), that in the long-time limit the agents are equally
distributed between the two internal states, $\theta \in \{+1,-1\}$. Hence,
we assume that on any location along the connection there are
$n_{\theta}= \bar{n}/2$ agents in state $\theta$.  Using this lower
limit for the agent number, the delta function in \eqn{hx-stat} can be
replaced by $\bar{n}/2=N/2A$ in the continuous limit.

Further, we consider that the agents move along the $x$ coordinate with a
constant velocity (which also matches with the assumption $\dot{v}\approx
0$ of the overdamped \name{Langevin} equation):
\begin{equation}
  \label{vx0}
  v=|\bbox{v}|=\frac{|x|}{t}\;;\;\; 0\leq x \leq L
\end{equation}
This allows us to replace $t$ in the time dependent production rate,
$s_{i}(\theta,t)$. With these simplified assumptions and the conventions,
\eqn{convent}, \eqn{hx-stat} reads for component $\theta=+1$:
\begin{equation}
  \label{hx-x}
  \frac{\partial h_{+1}(x,t)}{\partial t} = - k_{h}\,
  h_{+1}(x,t)+ \frac{\bar{n}}{2}\, s_{0} \exp\left\{
    -\beta\,\frac{x}{v}\right\} 
\end{equation}
 Integration of \eqn{hx-x} yields with $h_{+1}(x,t=0)=0$:
\begin{equation}
  \label{hx-t}
  h_{+1}(x,t)= \frac{\bar{n}}{2}\, \frac{s_{0}}{k_{h}}\, \exp\left\{
    -\beta\,\frac{x}{v}\right\} \Big(1\,-\,\exp\{-k_{h}\,t\}\Big) 
\end{equation}
Eventually, for $t\to\infty$ we find from \eqn{hx-t} the stationary
solution:
\begin{equation}
  \label{h1-stat}
  h_{+1}(x)= \frac{\bar{n}}{2}\frac{s_{0}}{k_{h}}\,
\exp\left\{-\frac{\beta}{v}\,x\right\}
\end{equation}
The remaining field component $h_{-1}(x',t)$ should have the same
stationary solution as \eqn{h1-stat}, with $x'=L-x$. The resulting total
field $\hat{h}(x,t)$ reads in the stationary limit (cf. \pic{h3}):
\begin{eqnarray}
  \label{htot-stat}
  \hat{h}(x) & = & h_{+1}(x)+h_{-1}(L-x)  \\
& =& \frac{\bar{n}}{2}\frac{s_{0}}{k_{h}}\,\left[\,
\exp\Big\{-\frac{\beta}{v}\,x\Big\}\,+\,
\exp\Big\{-\frac{\beta}{v}\,(L-x)\Big\}\,\right] \nonumber \\
&& 0\leq x \leq L \nonumber
\end{eqnarray}
\begin{figure}[htbp]
\centerline{\psfig{figure=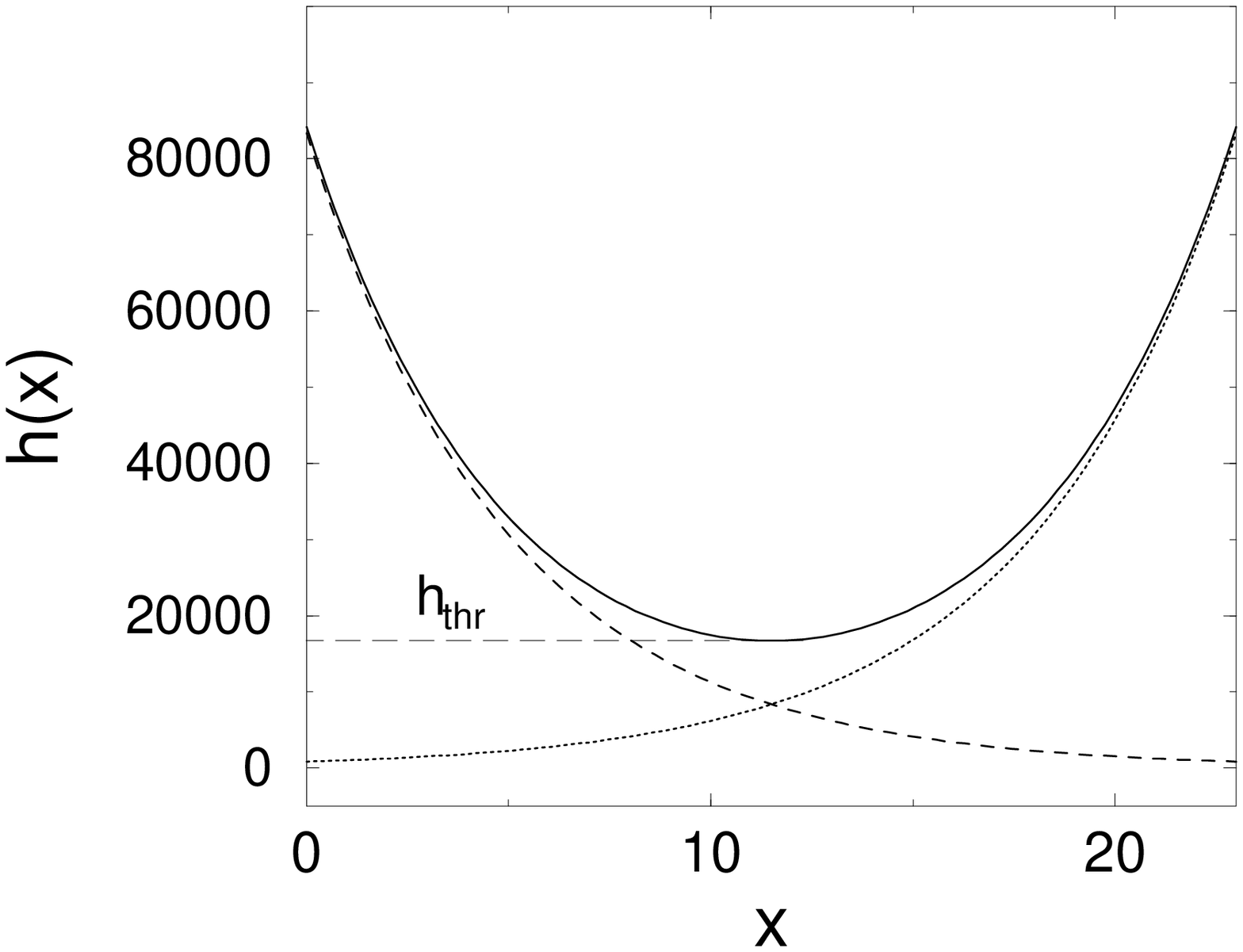,width=7.cm}}
\caption[p-h3]{
  Stationary solutions for $h_{+1}(x)$, \eqn{h1-stat} $(\cdots)$,
  $h_{-1}(x)$ $(--)$ and $\hat{h}(x)$, \eqn{htot-stat}
  $(-\!\!\!-\!\!\!-)$.  $L=L_{max}/2$, \eqn{l-max}. The threshold
  $h_{thr}$ is defined as the minimum value of $\hat{h}(\bbox{r},t)$ in
  the stationary limit.  For the parameters see \pic{t-rautpic}.
  \label{h3}}
\end{figure}

The threshold $h_{thr}$ should be defined as the minimum of
$\hat{h}(x)$, which yields for $L/2$. As the result, we find:
\begin{equation}
  \label{tres}
  h_{thr}=\hat{h}\left(\frac{L}{2}\right)= \frac{N}{A}
\frac{s_{0}}{k_{h}}\,\exp\left\{-\frac{\beta}{v}\,\frac{L}{2}\right\}
\end{equation}
Here, the threshold value is a function of the mean agent density
$\bar{n}$, the parameters $s_{0}$, $k_{h}$ and $\beta$ and the distance
between the two nodes, $L$.  However, due to the decay of the field
($k_{h}$) and the decreasing production rate with time ($\beta$), the
agents are only able to link nodes that are in a distance closer than a
critical distance $L^{\star}$, i.e. \eqn{tres} makes sense only for
$L<L^{\star}$.

In order to get an estimation for $L^{\star}$, we assume that a minimum
production rate $s_{min}$ exists, which is in the given model the
smallest possible amount of chemical released by the agent (naturally, it
could be a molecule, if $s$ is measured in molecule numbers). With
$t_{0}$ being the time when the agent hits the node, we get from
\begin{equation}
  \label{q-min}
  s_{min}=s_{0}\, \exp\Big\{-\beta (t-t_{0})\Big\}
\end{equation}
the maximum time $t_{max}$ after which the production is
\emph{negligible}:
\begin{equation}
  \label{t-max}
  t_{max}= \frac{1}{\beta} \ln\left\{\frac{s_{0}}{s_{min}}\right\}
\end{equation}
We can now discuss the case, that the agent moves straight with a
constant velocity, without changing its direction. Then the maximum
distance crossed before the contribution to the field is negligible,
would be:
\begin{equation}
  \label{l-max}
 L_{max}= v\, t_{max} =
\frac{v}{\beta} \ln\left\{\frac{s_{0}}{s_{min}}\right\}
\end{equation}
Contrary, if we assume that the agent moves like a random walker, the
average distance reached after $t$ simulation steps, is given by the mean
displacement, $\Delta R= \sqrt{2d\,D_{n}\,t}$, which yields for $d=2$:
\begin{equation}
  \label{displ}
  L_{av}= \sqrt{2\,D_{n}\,t_{max}}=\sqrt{\frac{2\,D_{n}}{\beta} \ln
\left\{\frac{s_{0}}{s_{min}}\right\}}
\end{equation}
The real maximum distance that can be connected by one agent in the
considered model, is of course between these limits. We have found 
\citep{fs-tilch-02-c} that $L_{max}/2$ is a reasonable estimate for
$L^{\star}$:
\begin{equation}
  \label{l-real}
  \sqrt{\frac{2D_{n}}{v}L_{max}}\,<\, L^{\star} 
\approx \frac{L_{max}}{2} \,<\, L_{max}
\end{equation}
Using this approximation, we find with \eqn{tres} and \eqn{l-max}
eventually the estimate for the threshold:
\begin{equation}
  \label{tres-f}
  h_{thr}= \frac{N}{A} \frac{s_{0}}{k_{h}}\,
\left(\frac{s_{min}}{s_{0}}\right)^{1/4}
\end{equation}
Provided the set of parameters used for the simulations, we find for the
threshold the value $h_{thr}= 1.7 \times 10^{4}$, which is approximately
$h_{thr}\approx 2 s_{0}$. We note again, that this is an
estimate that might give a rather high value because of the assumed
worse conditions.  On the other hand, it ensures that values for
$\hat{h}(a,t)$ above the threshold \emph{really} represent a
\emph{stable} connection $a$.
 
\subsection{Results of Computer Simulations}
After these theoretical considerations, we are now able to calculate the
connectivity $E$, \eqn{connec-e}, for the network simulated in
\pic{t-rautpic}. \pic{e-t} shows the increase of the connectivity in the
course of time.  In agreement with the visible evolution of the network
presented in \pic{t-rautpic}, three different stages can be clearly
distinguished:
\begin{enumerate}
\item an \emph{initial} period ($t<10^{2}$), where no
connections yet exist,
\item a \emph{transient} period
($10^{2}<t<10^{4}$), where the network establishes, 
\item a
\emph{saturation} period ($t>10^{4}$), where almost all nodes are
connected, and only small fluctuations in the connectivity occur.
\end{enumerate}
\begin{figure}[ht]
\centerline{\psfig{figure=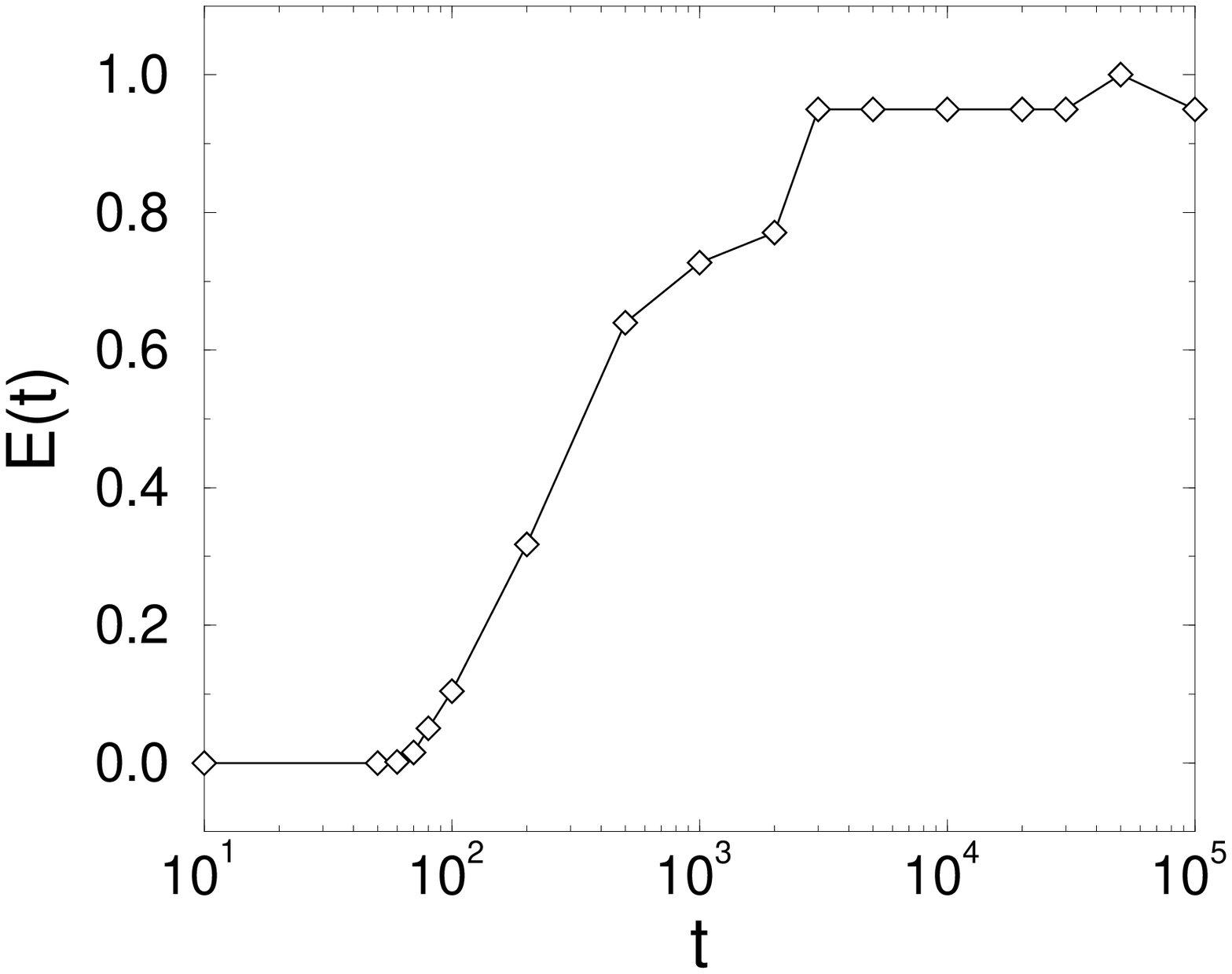,width=7.cm}}
\caption[]{
  Network connectivity $E$ \eqn{connec-e} vs. time $t$ in simulation
  steps, calculated from the series of  \pic{t-rautpic}. \label{e-t}
}
\end{figure}

\pic{e-t} results from the single realization of the network shown in
\pic{t-rautpic}, in the average we find certain fluctuations in the
connectivity due to stochastic influences that affect the formation of
the network during the transient period.  This leads us to the question
on what parameters the connectivity of the network depends. There are of
course the parameters affecting the production and decay of the two
different chemical cues, $s_{0}/k_{h}, \beta$, and thus the positional
information available to the agents. Another important parameter is the
average agent density $\bar{n}=N/A$.  If it is too low, the links will
not be established properly, either because not all nodes have been
detected during the transient period or because there are not enough
agents to maintain the links sufficiently. \pic{e-w} shows the average
connectivity $\mean{E}$ dependent on the density of agents, $\bar{n}$.
\begin{figure}[htbp]
\centerline{\psfig{figure=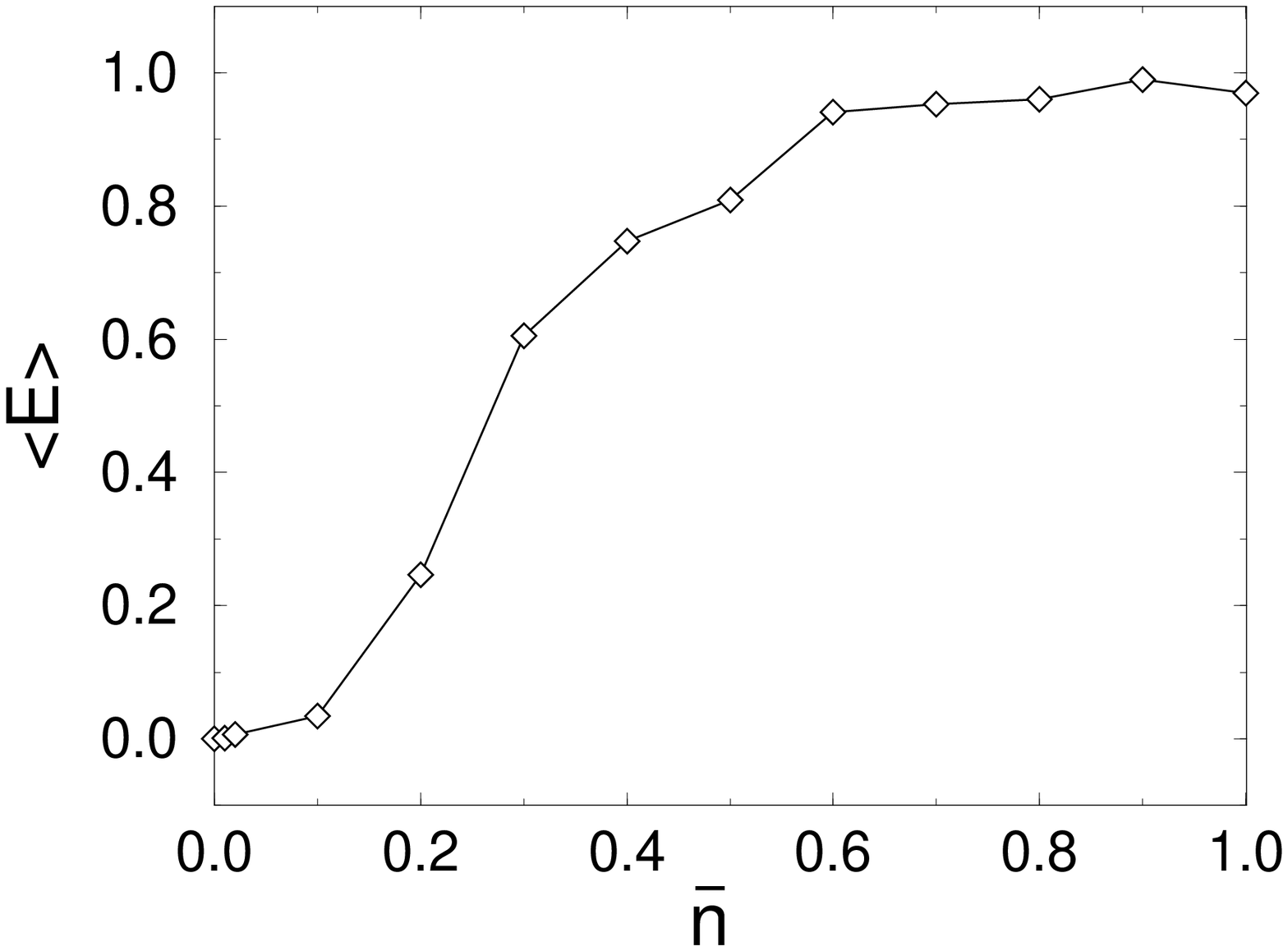,width=7cm}}
\caption[p-e-w]{
  Network connectivity $\mean{E}$ \eqn{connec-e}, averaged over $5$
  simulations vs. mean density of agents, $\bar{n}$.  Further parameters
  see \pic{t-rautpic}. \label{e-w}}
\end{figure}

Here, we clearly see that below a critical density the connectivity is
almost zero because not enough agents are available to establish the
connections. On the other hand, above a certain density the connectivity
reaches a saturation value that could be also below $1$, as \pic{e-w}
shows. Hence, at this point an increase of the number of agents does not
necessarily results in the establishment of more links. This is caused by
the \emph{screening effect} already mentioned in Sect. III, which
eventually concentrates all agents to move along the established links.

A third important impact on the establishment of the network results from
parameter relation between $\alpha_{i}$ and $\eps_{i}$ which influence
the agent's motion according to \eqn{langev-red2}.  If both the response
to the field, $\alpha_{i}$, and the sensitivity $\omega_{i}\propto
1/\eps_{i}$ are low, the agent nearly behaves as a random particle. On
the other hand, a strong response or a high sensitivity may result in a
decrease of stochastic influences, and the agent pays more attention to
the effective field that guides its motion. This in turn increases the
screening effect and may prevent the agent from discovering unknown
nodes, thus we can expect an optimal range of these parameters for an
efficient network formation. This has been investigated in more detail in
a subsequent paper \citep{fs-tilch-02-b}.

\section{Conclusions}

In this paper, we have proposed an agent-based model that shows the
\emph{self-assembling} of network-like structures between arbitrary
nodes.  Different from network models that start with the assumtion of a
known set of nodes to be linked in a straightforward manner, 
we have addressed in our model the question of how to connect a set of
nodes \emph{without} using prior information about their spatial
locations or preexisting long-range attraction forces.  This would need
to solve two problems in parallel: (i) the detection of the appropriate
nodes, and (ii) the establishment of one (or many) stable links between
them. 

As we have noted in the introduction, this is a scientific problem of
relevance in different areas, including the emergence of neural
connections. In the latter case, positional information in terms of
chemical gradients plays an important role in guiding the neural axons to
their destinations.  In our model, we have assumed that the positional
information about the existence of the nodes does not preexist, but is
generated ``on the fly'' by means of Brownian agents while they are
moving on the surface. Due to the non-linear feedback between the detected
positional information and the creation of new one, we find the emergence
of links between different nodes, along which the Brownian agents perform
a directed motion -- this way reinforcing and ``maintaining'' the links.

Different from a \emph{circuitry} for instance, for which the links
between the nodes are determined in a \emph{top-down} approach of
hierarchical planning, the connections here are created by the agents
\emph{bottom-up}, in a process of self-organization. As the computer
simulations have shown, the model turns out to be very flexible regarding
the geometry of the nodes to be connected.  Further, the networks created
this way are rather \emph{robust} against disturbances.  If for example a
particular link breaks down, the agents would be able to \emph{repair} it
by re-establishing the field, or by creating a new one.

The basic feedback mechanism in our model is given by the agent's
generation of (two) different kind of chemical information and the
agent's response to gradients of these chemicals dependent on their
internal state.  This is also known as \emph{chemotaxis}, i.e. the
response to (gradients of) chemical substances and is widely found on
different levels of biological organization. However, different from
other self-wiring models \citep{segev-physlett-98, segev-00} we have not
assumed that these substances due to their diffusion may have a
long-range effect on the moving agents, or may generate attractive
\emph{and} repulsive forces.  In our model, the chemical information acts
only locally.  It is stored in an effective field that is sometimes also
denoted as a \emph{self-consistent field}, because it is generated by the
agents and at the same time also influences their further behavior.  From
a more general perspective, the effective field plays the role of a
communication medium among the agents, i.e. it stores information
(external to the agents) for a certain time (determined by the decay rate
$k_{h}$) and allows access to it under certain conditions - for instance,
in the current model an agent can only access information that is at its
current position and is labeled differently from the value of the agent's
internal parameter. But the agents do not just respond to the information
provided in a purely reactive manner, they also actively change it
dependent on their internal parameter.

The Brownian agent concept has proven its use in a number of applications
where (positive and negative) local feedback processes play a
considerable role, but an internal \emph{evolution} of the agents can be
neglected  
\citep{fs-book-01}. The concept does not deny its inspiration from
statistical physics, using e.g. generalized Langevin equations for the
agents or reaction-diffusion equations for the effective field. Moreover,
it purposefully stretches these analogies, in order to apply methods from
statistical physics to derive pieces for a formal approach to
\emph{multi-agent systems} (MAS). This should also include the derivation
of a macroscopic dynamics of the MAS based on the agents
(\emph{microscopic}) dynamics, to allow some predictions of the
collective behavior and the derivation of critical parameters, etc.

On the first glimpse, agent-based approaches seem to be outside of the
realms of physics. Therefore, at the end we would like to comment on
this. First of all, ``agent'' or particle-based models are also useful in
physics if continuous approximations are not appropriate, for example in
cases where only small particle numbers govern the process (e.g. in
dielectrical breakdown or filamentary pattern formation).  Here
deterministic approaches or mean-field equations are not sufficient to
describe the behavior of the system, because the influence of history,
i.e.  stochastic fluctuations, early symmetry breaks, path dependence,
etc. play a considerable role -- which can be appropriately captured in
particle-based models. Moreover, these models also provide a very efficient
way to simulate structure formation processes by solving a large number
of coupled ``particle'' equations (such as Langevin equations) instead of
integrating a complicated set of coupled partial differential equations
\citep{lsg-mieth-rose-malch-95, lsg-fs-mieth-97}

In many interdisciplinary applications physics nowadays deals with, for
example in biological physics or econophysics, the basic system entities
do not just respond to interaction forces, but also perform certain types
of activities, such as active motion or changes of the ``environment''
and further have internal degrees of freedom that allow them to act
differently. Therefore, the ``physical'' particle-based approach has been
extended towards an \emph{agent-based approach}, where the agents already
have an intermediate complexity to allow for certain non-trivial actions
or responses.  In the model discussed in this paper, the generation of
two different chemical cues and the complex response to them are just two
examples for such an extension.  In addition to the Browian agent model,
other agent-based approaches have been developed based on physical
principles, which focus on particular aspects of the complex agent
behavior.  We just mention here \emph{active walker models}
\citep{kayser-et-92,fs-lsg-94,lam-95,helbing-fs-et-97,sheu-et-99}, where
a potential can be locally changed by the walker, or \emph{active
  Brownian particles} \citep{lsg-mieth-rose-malch-95, fs-97-agent} that
deal also with the energetic aspects of active motion.  With respect to
specific biological phenomena, there are the \emph{communicating walker
  model} used in the study of complex patterning of bacterial colonies
\citep{ben-jacob-et-94-a}, or the \emph{bions model} used in the study of
amoebae aggregation \citep{kessler-levine-93}, or the many models of
self-driven particles to describe swarming behavior
\citep{mikhailov-zanette-99, czirok-vicsek-00, levine-00,
  fs-eb-tilch-01}.  They jointly demonstrate, that agent-based models can
indeed profit from the methodology and the tools derived in statistical
physics.


\begin{acknowledgments}
  The authors would like to thank L. Schimansky-Geier (Berlin) for
  discussions. This manuscript was completed during a stay at the Centro
  de Ciencias Matematicas (CCM) of the Universidade da Madeira
  (Portugal). F.S. would like to thank L. Streit for the kind
  hospitality.
\end{acknowledgments}

\bibliography{net-submit}
\end{document}